# Signal-to-noise per unit time optimization for *in vivo* single-voxel proton magnetic resonance spectroscopy of the brain: Theoretical formulation and experimental verification at two field strengths

D.E. Gkotsis[a,b], E.D. Gotsis[c], E.P. Pappas[a], E.Z. Kapsalaki[b,d], and I. Seimenis[a]

[a] National and Kapodistrian University of Athens, Medical School, Department of Medical Physics, Athens, Greece

[b] Institute Euromedica–Encephalos, Magnetic Resonance Department, Athens, Greece

[c] Institute Bioiatriki, Magnetic Resonance Department, Athens, Greece

[d] University of Thessaly, Medical School, Radiology Department, Larissa, Greece

**Corresponding author:**

D.E. Gkotsis, PhD

Department of Medical Physics

School of Medicine

National and Kapodistrian University of Athens

75 Mikras Asias Street

Goudi, 115 27

Athens, Greece

Phone (Mobile): (+30) 6972 458 156

Email: d.gkotsis@med.uoa.gr

**Signal-to-noise per unit time optimization of *in vivo* single-voxel proton magnetic resonance spectroscopy in brain oncology: Theoretical formulation and experimental verification at two field strengths**

**Abstract**

Signal-to-noise ratio optimization, regarding repetition time selection, was explored mathematically and experimentally for single-voxel proton magnetic resonance spectroscopy. Theoretical findings were benchmarked against phantom measurements at 1.5 Tesla and localized *in vivo* proton brain spectra acquired at both 1.5 Tesla/3.0 Tesla. A detailed mathematical description of signal-to-noise ratio per unit time was derived, yielding an optimal repetition time of 1.256 times the metabolite longitudinal relaxation time. While long-repetition-time acquisitions minimize longitudinal relaxation time contributions, a repetition time of ~1.5s results in maximum signal-to-noise ratio per unit time, which can in turn be invested into smaller voxel sizes. The latter is of utmost importance in brain oncology, allowing accurate spectroscopic characterization of small lesions.

**Introduction**

In *in vivo* magnetic resonance imaging (MRI) and magnetic resonance spectroscopy (MRS), acquisition time is one of the most important parameters to consider from both the patient's and health care provider's perspectives. Similarly, signal to noise ratio (SNR) is one of the most significant image (or spectral) quality indices. A high SNR characterizes a system of good overall performance, allowing the identification of challenging (due to small size) lesions.

SNR depends on several factors. For example, it is proportional to the volume of the voxel employed, and the square root of the number of signal averages (excitations) performed. Since averaging takes time, attained SNR is closely related to the total acquisition time. Additionally, an increase in the repetition time (TR) increases the SNR since a long TR allows the longitudinal magnetization to approach its maximum, thus producing high signal intensities. Therefore, the choice of proper acquisition parameters to achieve the highest SNR possible for a given total acquisition time is unquestionably important.

The question of maximizing the SNR per unit time was initially dealt with by Ernst (1, 2) as early as 1966, a few decades prior to the translation of MRS in the clinical practice. Ernst's famous solution $\cos(\theta) = e^{-\frac{TR}{T1}}$, or equivalently $\theta = \arccos\left(e^{-\frac{TR}{T1}}\right)$, defines the flip angle, θ, of the excitation radiofrequency (RF) pulse to provide the highest SNR for a given longitudinal relaxation time, $T_1$.

However, in *in vivo* single voxel MRS, as performed in routine clinical practice with commercially available (i.e., no research-mode) MR scanners and pulse sequences, spatial localization is achieved by volume-selective echo techniques which utilize a 90º pulse to initially generate transverse magnetization (3) and, therefore, the practitioner cannot employ pulses of variable flip angle to maximize SNR. This limitation, therefore, imposes some constraints on the choice of acquisition parameters, which in turn affect the SNR that can be achieved in a given amount of time and, ultimately, the quality of the acquired data.

In 1970, J.S. Waugh (4) published a paper dealing with the sensitivity in Fourier Transform NMR spectroscopy, quoting a value of $TR/T_1 = 1.269$ for maximum SNR using 90º RF pulses.

Becker et al (5) and Hendrick et al (6) also provided qualitative descriptions on the subject, wherein a detailed mathematical derivation of the optimization problem was not demonstrated.

An extended literature review did not reveal a detailed mathematical solution for the maximum SNR as a function of TR for 90º RF pulses, and this served as our initial motivation for proceeding with the derivation of a relevant theoretical treatment.

Time is particularly important in *in vivo* $^1$H-MRS, since a large number of excitations (NEX) is usually employed to improve SNR. However, the same acquisition time can be achieved using different combinations of TR and NEX. The two main spectroscopic techniques most commonly available in commercial MR scanners are stimulated echo acquisition mode (STEAM), and point resolved spectroscopy (PRESS), both of which are based on a saturation recovery scheme.

Typically, saturation recovery sequences employ large TR values to maximize the signal. For example, Wilken et al proposed a TR of 6s to minimize $T_1$ effects on the spectra (7), whilst several investigators reported TRs between 5 and 6 seconds (8-11).

This work is motivated by the potentially suboptimal adjustment of TR for the acquisition of single voxel MRS data in the clinical setting, and specifically for brain oncology spectroscopic acquisitions. It aims, therefore, at reconfirming previous qualitative suggestions (4-6) regarding the proper adjustment of TR for acquiring spectra with optimal SNR per unit time and at directing users' attention to this issue. To this scope, the signal to noise differential equation is solved analytically and a formalism is provided. An experimental study and two *in vivo* brain measurements (at 1.5T and 3.0T) were also performed to verify the theoretical findings.

**Materials and Methods**

The study involving human participants (1 healthy volunteer for visualization purposes) was exempted from the institutional review board, as it was one of the authors (DEG), who underwent the spectroscopic brain acquisitions presented in Figure 4, who also consented to presenting the acquired spectra herein, and the data acquisition involved neither use of ionizing radiation, nor excessive scan times.

Theoretical part

The NMR signal equations can be derived directly from the Bloch equations, which are known since the early days of NMR and, for gradient echo pulse sequences, they can be summarized in equation (1):

$$S(\theta, TR, TE) = S_0 \sin\theta \frac{1-e^{-TR/T_1}}{1-\cos\theta \cdot e^{-TR/T_1}} e^{-TE/T_2^*} \qquad (1)$$

TE is the echo time of the pulse sequence, $S_0$ denotes the spin density, and $T_1$ and $T_2$ are the longitudinal and transverse relaxation times, respectively.

By substituting $T_2^*$ with $T_2$, and considering the limit where the flip angle tends to 90º, equation (1) then reduces to the typical saturation recovery signal equation:

$$S(TR) = S_0\left(1 - e^{-TR/T_1}\right)e^{-TE/T_2} \tag{2}$$

where $S_0$ represents the magnetization at equilibrium (or equivalently, the spin density).

Since thermal noise (i.e., noise due to thermal fluctuations of the patient spins and in the system's electronics) is random (white), while the signal is coherent, addition of the acquired signals results in a signal to noise ratio improvement proportional to the square root of NEX. Thus, equation (2) can be modified to consider more general cases where NEX > 1, yielding an expression for SNR:

$$SNR(TR) \sim S_0(1 - e^{-\frac{TR}{T_1}})e^{-TE/T_2}\sqrt{NEX} \tag{3}$$

NEX can be expressed as the total acquisition time (TAT) divided by TR:

$$NEX = TAT/TR \tag{4}$$

Considering the clinically relevant MR acquisition scenario of a given TAT (i.e., time available to perform data acquisition), the described analysis can be further manipulated by substituting equation (4) into equation (3) and treating TAT as a constant. Thus, TAT can be incorporated into $S_0$. Finally, since the TE is constant throughout the acquisition, the effect of $T_2$ relaxation can also be regarded as constant for any TE used. By denoting this proportionality constant ($S_0 e^{-TE/T_2}\sqrt{TAT}$) as k, equation (2) simplifies to:

$$SNR(TR) = k(1 - e^{-TR/T_1})/\sqrt{TR} \tag{5}$$

Then, the possible maxima or minima of the function must be determined by evaluating the first and second derivatives of SNR(TR) with respect to TR. Substituting t for the TR in equation (5), the following expression is obtained:

$$S(t) = \frac{k\left(1-e^{-\frac{t}{T_1}}\right)}{\sqrt{t}} \quad (6)$$

The first and second derivatives, respectively, of S(t) with respect to t are:

$$\frac{dS(t)}{dt} = k\left[\frac{1}{T_1}\frac{1}{\sqrt{t}}e^{-t/T_1} - \frac{1}{2}\frac{(1-e^{-t/T_1})}{t^{3/2}}\right] \quad (7a)$$

and

$$\frac{dS^2(t)}{dt^2} = -k\left[\frac{1}{T_1^2}\frac{1}{\sqrt{t}}e^{-t/T_1} - \frac{1}{T_1}\frac{e^{-t/T_1}}{t^{3/2}} + \frac{3}{4}\frac{(12e^{-t/T_1})}{t^{5/2}}\right] \quad (7b)$$

Analysis of equation (7b) shows that it is negative, thus the original function has a maximum. That maximum occurs at the value at which the first derivative is zero:

$$\frac{dS(t)}{dt} = k\left[\frac{1}{T_1}\frac{1}{\sqrt{t}}e^{-t/T_1} - \frac{1}{2}\frac{(1-e^{-t/T_1})}{t^{3/2}}\right] = 0 \quad (8)$$

Therefore, the equation to be solved is the following:

$$\left[\frac{1}{T_1}\frac{1}{\sqrt{t}}e^{-t/T_1} - \frac{1}{2}\frac{(1-e^{-t/T_1})}{t^{3/2}}\right] = 0 \quad (9)$$

Experimental part

To experimentally test the validity of equation (5), a suitable phantom consisting of 15 vials was constructed. Each vial, having a square cross section of 2 cm x 2 cm and a height of 10 cm, was filled in with an aqueous $MnCl_2$ solution of variable concentration (ranging from 0 to 3.2 mM). The phantom with the 15 solutions of different $T_1$ relaxation times was scanned at 1.5 Tesla (Signa HDxt, GE Healthcare, Milwaukee, US). The phantom SNR measurements as a function of TR were performed using a constant TAT. The product NEX×TR and, thus, the TAT were maintained constant by appropriately manipulating both NEX and TR. Specifically, a total acquisition time TAT = 1920 msec can be achieved by employing different combinations of TR×NEX i.e., TR = 1920 ms and NEX = 1, or TR = 960 ms and NEX = 2, or TR = 480 and NEX = 4, and so on and so forth.

To measure the phantom $T_1$ values, an initial measurement with a partial saturation recovery sequence gave a first approximation of the $T_1$ for the solution in each vial. Appropriate saturation recovery experiments were then designed to accurately measure the low $T_1$ values of the phantom (using a TR range of 20 to 600 ms), the intermediate $T_1$ values of the phantom (using a TR range of 30 to 1200 ms), as well as the high $T_1$ values of the phantom (using a TR range of 60 to 1920 ms). This approach was adopted to shorten the overall scanning time, while ensuring that the highest TR value used in each set was, at least, higher than 2.5 times the $T_1$ of each sample in the set (although this condition does not stand true for the pure water with $T_1 > 3500$ ms). The constant NEX and variable TRs employed in this scheme resulted in variable TAT for each saturation recovery experiment.

To validate theoretical findings with *in vivo* measurements, two single-voxel spectra were acquired at 1.5T (GE Signa HD28) and two more at 3.0T (GE Signa Premier Air$^{TM}$), from the brain of one of the authors (DEG), who gave consent for presenting relevant data in this study. All spectra were obtained from a 1.35 x 2.05 x 3.82 cm$^3$ voxel located in the left centrum semiovale white matter, while the two acquisitions per field strength were consecutive. The first acquisition was performed with a TR of 1.5 s and 64 averages, whilst the second one employed a TR of 6.0 s and 16 averages, and all other acquisition parameters were identical. Thus, acquisition time of any water-suppressed spectra was 96 s.

However, total scan time for the long TR measurements was almost double that for short TR (240 s vs 132 s), since the corresponding water-unsuppressed acquisition (used for artifact reduction, enhanced water suppression, etc.) incorporated in all MRS protocols employs a constant number of averages (24), regardless of the total number of averages chosen for the water-suppressed acquisition. Therefore, the time duration prolongation of the water-unsuppressed spectra is directly influenced by TR selection.

Data post-processing and statistical analysis

The SNR calculations of the multi-vial phantom were based on the following equation:

$$SNR = 0.655 \frac{S}{\sigma} \qquad (10)$$

where S is the average signal intensity within each vial and σ is the standard deviation of the background noise measured in a large region of interest adjacent to the phantom (i.e., in air free from signal or artefacts). The 0.655 factor corrects for the fact that the noise in a magnitude MR image does not follow a Gaussian distribution with zero mean (12,13). Specifically, Edelstein et al. showed in 1984 (12) that pure noise in magnitude images is governed by the Rayleigh distribution, whilst Bernstein et al. in 1989 (13) provided the closed form solution of the more general Rician distribution in their study on detectability in MRI. Both S and σ were measured offline using the RadiAnt software (Medixant RadiAnt DICOM Viewer, URL: https://www.radiantviewer.com). The calculated SNR values were fitted to equation (5) using a least square fitting routine in EXCEL (Microsoft, Redmond, US) to also obtain phantom $T_1$ values. $T_1$ values acquired from equation (5) were benchmarked against the corresponding values provided by the saturation recovery method:

$$S = So\,(1 - e^{-TR/T_1}) \qquad (11)$$

with S being the average signal intensity within each vial and $S_0$ being the initial magnetization.

Plots to compare the $T_1$ values estimated using the SNR measurements against those provided by the conventional SR method, as well as statistical analyses, were carried out with the MedCalc Statistical Software version 19.4.1 (MedCalc Software Ltd, Ostend, Belgium).

**Results**

Theoretical part

*Solution 1*

Equation (9) can be rearranged to:

$$-\frac{1}{2t^{3/2}} + \frac{e^{-t/T_1}}{2t^{3/2}} + \frac{e^{-t/T_1}}{T_1} \cdot \frac{1}{t^{1/2}} = 0 \tag{12a}$$

Equation (12a) is equivalent to:

$$e^{-t/T_1} \cdot \left(\frac{1}{t^{1/2}T_1} + \frac{1}{2t^{3/2}}\right) = \frac{1}{2t^{3/2}} \tag{12b}$$

Further manipulation of the left side term leads to:

$$e^{-t/T_1} \cdot \left(\frac{2t+T_1}{2t^{3/2}T_1}\right) = \frac{1}{2t^{3/2}} \Rightarrow e^{-t/T_1} = \frac{\frac{1}{2t^{3/2}}}{\frac{2t+T_1}{2t^{3/2}T_1}} \tag{12c}$$

which results in:

$$e^{-t/T_1} = \frac{T_1}{2t+T_1} \tag{13}$$

Equation (13) does not have an exact closed form solution, but it can be solved graphically. Figure 1(a) depicts the graphical solution of equation (13).

*Solution 2*

By rearranging the terms of equation (9) differently, and by multiplying both sides of the resulting equation with the term *($e^{t/2T_1}$)$t^{3/2}$*, we obtain:

$$\frac{(e^{t/2T_1}-e^{-t/2T_1})}{2} = \frac{t}{T_1}e^{-t/2T_1} \qquad (14)$$

The left side is the hyperbolic sine function, sinh(t/2$T_1$), thus equation (14) becomes:

$$sinh(t/2T_1) = \frac{t}{T_1}e^{-t/2T_1} \qquad (15)$$

Solutions of equation (15) are provided by Lambert's W function with t/$T_1$ = 0 and t/$T_1$ = 1.256. Figure 1(b) represents the graphical solution of equation (15).

Although equations (13) and (15) appear to be different in form, they do have identical solutions, as shown in Figures 1(a) and 1(b). Rejecting the trivial solution at t/$T_1$ = TR/$T_1$ = 0, the solution at which the SNR per unit time is maximum occurs for TR/$T_1$ = 1.256. It is therefore evident that the maximum SNR per unit time is achieved at TR = 1.256×$T_1$, as confirmed by the graphical and analytical solutions of equation (9).

Experimental part

Figure 2 depicts three representative least squares fits of equation (5) to the SNR data of the multi-vial phantom acquired at variable TRs but at constant total acquisition time. Least squares fitting allowed the estimation of the $T_1$ value for each vial (e.g., 98.1 ms, 224.4 ms and 713.7 ms for 3 vials whose data fitting is presented in Figures 2a, 2b and 2c, respectively). A scatter plot between the $T_1$ values derived by the least squares fitting of equation (5) against the corresponding values provided by the SR method is depicted in Figure 3a. Linear regression analysis revealed an excellent correlation between the two datasets ($R^2=0.9998$). In addition, a paired t-test performed on the data comprising the set of measurements on the multi-vial phantom manifested no statistically significant differences between the two approaches. Furthermore, a Bland-Altman analysis (not shown) did not reveal any systematic differences between the two methodologies.

As demonstrated in Figure 3b, a good linear correlation ($R^2=0.9828$) was found between the relaxation rate $R_1$ ($=1/T_1$) provided by the proposed method and the concentration of the paramagnetic ions, although a slight deviation was noted at relatively high concentrations due to $T_2$ effects (the use of a gadolinium chelate would have eliminated this effect). An even better correlation ($R^2=0.9946$) was observed between $R_1$ values derived by fitting saturation recovery measurements to equation (5) and paramagnetic ion concentrations.

Figure 4 displays the left centrum semiovale water-suppressed spectra acquired at 1.5T (A and B) and at 3.0T (C and D) from the brain of one of the authors (DEG), with the two different TR/NEX combinations employed but at the same TAT. At both field strengths, SNR values of all metabolites were higher in the spectra acquired with TR/NEX = 1500 ms/64 (A, C) compared to those obtained from the TR/NEX = 6000 ms/16 spectra (B, D).

For creatine (Cr), which usually represents an internal reference marker allowing for metabolite ratio calculations, the SNR difference between the two spectra was 22% at 1.5T and 29% at 3.0T. For choline (Cho), a marker of cellular membrane turnover elevated in many pathological states, SNR increases were 40% and 45% at 1.5T and 3.0T, respectively. The SNR of N-Acetyl-L-aspartate (NAA), a marker of neuronal viability and integrity, was found to increase by 22% at 1.5T and by 34% at 3.0T. As expected, for the same acquisition scheme, all metabolite SNRs were markedly elevated at 3.0T compared to 1.5T. Another key factor contributing to the overall SNR elevation at 3.0T was the use of a brain-dedicated, 48-channel, phased-array coil, whereas a more conventional, 16-channel head and neck phased-array coil was used at 1.5T.

**Discussion**

Although NMR Spectroscopy has been around for nearly 75 years, *in vivo* $^1$H-MR Spectroscopy has been used in clinical practice for about 30 years. Major improvements have been noted in the field during this period, but the exam remains challenging in the clinical setting. In most cases, there is a need for spectral acquisitions from more than one anatomical region, while $^1$H-MRS is often combined with a full MRI exam, thus posing severe restrictions on the afforded acquisition time per scan. *In vivo* spectroscopic sequences achieve localization of the region of interest with a combination of slice selective RF pulses (3). The most widely used spectroscopic sequences are STEAM (14) and PRESS (15). STEAM uses three slice-selective 90º RF pulses, and PRESS uses one 90º and two 180º slice selective pulses. It is important that accurately calibrated 90º or 90º/180º RF pulses are implemented, to achieve good localization (as well as maximum MR signal at the TR selected). STEAM holds several advantages compared to PRESS. A very short TE can be employed allowing the detection of metabolites exhibiting short transverse $T_2$ relaxation times, while at the same time, exclusive use of 90º pulses allows for better voxel edge definition and lower tissue energy deposition. STEAM sequence, however, yields about 50% the SNR of PRESS at the same TR's and TE's, since stimulated echoes, rather than spin echoes, are used. As a result, most investigators prefer PRESS over STEAM in the clinical setting.

All the above demonstrate the importance that the optimization of SNR per unit time holds in clinical *in vivo* MR spectroscopy. A review of the literature reveals brain studies in which the choice of acquisition parameters, and specifically of the repetition time, may be suboptimal regarding the contradictory needs between a relatively high SNR and an acceptable acquisition time (7-11). Those researchers used relatively long TR values to minimize $T_1$ contributions, resulting however in long acquisition times. To reduce acquisition time, many investigators have utilized a TR between 1700 and 4000 ms at 1.5 Tesla (16 – 19), whilst most of the preset protocols proposed by the scanner manufacturers employ a TR of around 2 seconds.

To reduce total acquisition time per voxel to an acceptable level, the main acquisition parameters affecting scan time, i.e., the repetition time TR and the number of excitations NEX, must be properly adjusted without however compromising the acquired spectrum quality in terms of SNR.

Therefore, a relationship between TR and $T_1$ that maximizes the SNR per unit time was derived and may find application in all routine spectroscopic acquisition approaches used in clinical spectroscopy. Mathematical analysis and experimental verification conducted in this work demonstrate that a $TR/T_1$ ratio of 1.256 maximizes the SNR for any given acquisition time. This result also reconfirms, for the purposes of localized *in vivo* MRS, the relevant qualitative findings made by Waugh (4) and Becker et al (5) (1.269 and 1.27, respectively) prior, however, to the development and implementation of *in vivo* MR Spectroscopy in the clinical setting.

As can be seen from Figure 1c, the signal intensity as a function of TR produces a plateau at about ± 15% of the optimal value $TR/T_1 = 1.256$. This means, essentially, that if the practitioner chooses the TR to lie in the range $1.15 \times T_1 \leq TR \leq 1.45 \times T_1$, the SNR per unit time achieved will also be optimal.

Let us consider, for example, the metabolite Creatine (Cr), with a $T_1$ value of around 1200 ms. The lower TR limit would be given by $0.85 \times 1.256 \times 1200$ ms, while the upper TR limit would be given by $1.15 \times 1.256 \times 1200$ ms. The lower limit means that the TR must definitively be larger than $0.85 \times 1.256 \times 1200$ ms, hence TR > 1300 ms (approx.), while the upper limit means that the TR must definitively be lower than $1.15 \times 1.256 \times 1200$ ms, hence TR < 1750 ms. This means that, essentially, if the practitioner selects the parameter TR to be in the range 1300 ms < TR < 1750 ms, maximum SNR per unit time will be achieved.

As demonstrated by data presented in Figure 4, simply increasing the TR from 1.5 s to 6 s decreases the Cho/Cr ratio by 13% and 11% at 1.5T and 3.0T, respectively, due to the greater effect of TR choice on the Cho peak compared to the Cr peak. Notwithstanding noticeable variations which are dependent on the brain region from which spectra were acquired, it has been shown that the brain Cho and Cr $T_1$ values hover at around 1200 ms at both 1.5T and 3T, while NAA and myoinositol (mI) exhibit slightly longer and shorter $T_1$ values, respectively (20). Thus, a TR of about 1500 ms may be regarded as a reasonable choice with respect to the SNRs achieved for the main metabolites considered in brain $^1$H-MRS. Regarding the influence of static magnetic field intensity on the $T_1$ relaxation times of brain metabolites, the evidence is conflicting (20, 21). Data derived from this study are suggestive of a moderate influence of the field strength on the centrum semiovale $T_1$ values, since SNR enhancements between 1.5 s and 6 s TRs realized at 3.0T are different than those reached at 1.5T. In this respect, the suggested TR/$T_1$ ratio may also find application in contemporary brain MRS studies performed at ultra-high field strengths (i.e., 7.0T), where a prolongation in the $T_1$ values of brain metabolites is anticipated. Although 7.0T systems inherently increase SNR, optimal acquisition parameter selection for SNR optimization remains an issue (22).

An *in vivo* $^1$H-MR spectrum usually consists of signals from more than one spin species which can exhibit different longitudinal relaxation times. Strictly speaking, therefore, the optimized TR value may only apply to one selected metabolite signal in the spectrum, with the other metabolite signals exhibiting a suboptimal signal intensity. This situation may have an impact on the estimated ratios between different metabolites, and in extracting absolute metabolite concentrations, both of which are often used as indicators for disease diagnosis, differentiation, and grading, as well as for monitoring disease progression and evaluating treatment efficiency. In most oncological patients who are undergoing *in vivo* MRS for the purposes of an accurate differential diagnosis, however, most metabolites are evaluated by the neuroradiologists in an all-or-none basis.

For instance, in differentiating between a grade IV glioblastoma multiforme and a typical metastatic grade IV brain tumor, from the point of view of $^1$H-MRS, the correct diagnosis depends on the presence/absence of creatine (metastatic tumors do not have creatine whereas diffusive tumors do). Hence, it is demonstrably more important to be able to achieve high SNR, thus increasing the experiment's sensitivity by being able to detect small metabolite peaks (compared to the overwhelming $CH_2$ and $CH_3$ lipid peaks at 1.3 and 0.9 ppm, respectively, that are always present in grade IV tumors), than having the $T_1$ effects minimized and obtaining more accurate absolute metabolite quantification.

Furthermore, by opting for a TR of about 1.5 s, rather than long (2-3 s) or ultra-long (5-6 s) TRs, the clinical practitioner is essentially given the opportunity to acquire spectra from much smaller voxel sizes and, therefore, to accurately characterize small lesions, even down to 0.2 cm$^3$ in volume. For such small lesions, the clinical practitioner will need to employ more than 450 signal averages, depending both on the lesion's proximity to the surface coil elements (i.e., lesions located near the center of the brain suffer from lower signal received at the coil), and on the overall local extent of malignancy where the voxel is placed (i.e., grade III/IV brain tumors contain more cells than healthy brain parenchyma and hence more protons contribute to the overall signal within the voxel). Therefore, at such conditions, a choice of high or ultra-high TR will render the acquisition infeasible.

For example, a small voxel size of the order 0.2 – 0.3 cm$^3$, would require more than 460 NEX (at 3.0T), and perhaps even more than that at lower field strengths. By opting for an ultra-long TR = 6000 ms, the TAT would be given by the sum of the acquisition time corresponding to the water-suppressed spectra (in this case 6000 ms × 460) and the acquisition time corresponding to the water-unsuppressed spectra (for GE scanners 6000 ms × 24), yielding a TAT of approximately 50 minutes, which is infeasible in clinical practice. On the other hand, opting for the proposed TR = 1500 ms, yields a TAT of approximately 12 minutes, which is certainly feasible.

To sum up, while long TR acquisitions do not suffer from $T_1$ effects and are, therefore, more reliable in predicting metabolite absolute concentrations, they impose severe spatiotemporal restrictions which could lower the experiment's sensitivity and minimize its overall contribution to a differential diagnosis in patients with a known or suspected brain tumor.

$T_1$ values of vials acquired with the proposed time-efficient approach were only compared against corresponding saturation recovery results. A thorough presentation and use of available methods for measuring $T_1$ values with imaging systems was beyond the scope of this study. However, future studies may incorporate saturation recovery, inversion recovery, and multiple flip angle techniques, as well as the more recently presented (23) saturation-inversion-recovery sequence which seems to provide considerably sharper $T_1$-dependence compared to inversion recovery and saturation recovery sequences.

Our study focused on the optimization of SNR per unit time in single-voxel $^1$H-MRS acquisitions, with emphasis on oncological brain cases. Also, the study exclusively concentrated on TR selection and, consequently, on total scan time. In clinical $^1$H-MRS, however, selection of TE is crucial, particularly when data acquisition with only a single TE is feasible. To decide which TE is the most appropriate for a particular data acquisition, it is critical to consider the differences in metabolite intensities caused by the effects of $T_1$ and $T_2$ relaxation times at the field strength of measurement (24).

**Conclusions**

The current study confirms the strong effect that acquisition parameters may have on the quality of proton MRS spectra, as well as on the time required for their acquisition. In *in vivo* single-voxel $^1$H-MRS of the brain, total acquisition time per spectrum is usually determined by system availability, patient conformity and the exam protocol prescribed by diagnostic needs. If a TR equal to 1.256 times the $T_1$ of a specific metabolite is chosen, then the SNR of that metabolite peak will be optimized for any given acquisition time. In addition, users must be aware of the potential changes in relative metabolite peaks and in calculated metabolite ratios that a TR modification may effectuate.

The proposed experimental scheme (variable TR and NEX, with the product of TR×NEX being constant) pertaining to SR pulse sequences, may serve as an alternative way to estimate tissue $T_1$ values.

The proposed TR selection produces maximum SNR per unit time for most major metabolites commonly detected *in vivo* in the human brain, and in magnetic field strengths of up to 3.0T. The SNR-per-unit-time surplus produced from the proposed parametrization can be further invested into smaller voxel sizes (and higher number of signal averages), allowing the clinical practitioner to investigate smaller lesions in reasonable acquisition times, without having to resort to hindering partial volume averaging effects.

Considering the most common MR spectroscopic acquisition conditions encountered in the routine clinical practice for oncology patients, and by opting for a short TR of about 1500 ms (instead of long-TR or even ultra-long TR acquisitions), the practitioner manages to achieve maximum metabolite SNR per time (for magnetic field strengths up to 3.0T).

While some practitioners prefer to opt for ultra-long TR acquisitions due to minimization of $T_1$ contributions, and consequently due to more robust metabolite quantification (7-11), spatiotemporal restrictions (i.e., voxel size, acquisition time etc.) are imposed which directly influence the examination's overall sensitivity by disabling the practitioner from acquiring spectra from small lesions, which is of utmost importance in brain oncology (i.e., early-stage tumor characterization).

**Declarations**

*Declarations of interest*: None

*Conflicts of interest:* None

**Funding**

This work was supported, in part, by the State Scholarships Foundation (I.K.Y.) and has been co-funded by Greece and the European Union (European Social Fund) via the operational *program "Human Resources* Development, *Education* and *Lifelong Learning* 2014-2020".

Word count: *5623*

**References**

1. Ernst RR. Sensitivity enhancement in Magnetic Resonance. *Adv Magn Reson* **1966**; 2(1)

2. Ernst RR, Anderson WA. Application of Fourier transform spectroscopy to magnetic resonance. *Rev Sci Instrum* **1966**; 37(1): 93-102

3. Keevil SF. Spatial localization in nuclear magnetic resonance spectroscopy. *Phys Med Biol* **2006**; 51(16): 579-636

4. Waugh JS. Sensitivity in Fourier transform NMR spectroscopy of slowly relaxing systems. *Journal of Molecular Spectroscopy* **1970**; 35(2): 298-305

5. Becker ED, Ferretti JA, Gambhir PN. Selection of Optimum Parameters for Pulse Fourier Transform Nuclear Magnetic Resonance. *Analytical Chemistry* **1979**; 51(9): 1413-1420

6. Hendrick RE, Newman FD, Hendee WR. MR Imaging Technology: Maximizing the Signal-to-Noise Ratio from a Single Tissue. *Radiology* **1985**; 156: 749-752

7. Wilken B, Dechent P, Herms J et al. Quantitative Proton Magnetic Resonance Spectroscopy of Focal Brain Lesions. *Pediatr Neurol* **2000**; 23(1): 22-31

8. Brockmann K, Pouwels PJ, Dechent P, Flanigan KM, Frahm J, Hanefeld F. Cerebral proton magnetic resonance spectroscopy of a patient with giant axonal neuropathy. *Brain Dev* **2003**; 25(1): 45-50

9. Watanabe T, Frahm J, Michaelis T. Amide proton signals as pH indicator for in vivo MRS and MRI of the brain-Responses to hypercapnia and hypothermia. *Neuroimage* **2016** Jun; 133: 390-398

10. van der Graaf M. In vivo magnetic resonance spectroscopy: basic methodology and clinical applications. *Eur Biophys J* **2010**; 39(4): 527-40
21

**Figures**

*Graphical confirmation of the optimal repetition time TR selection as a function of $T_1$.*

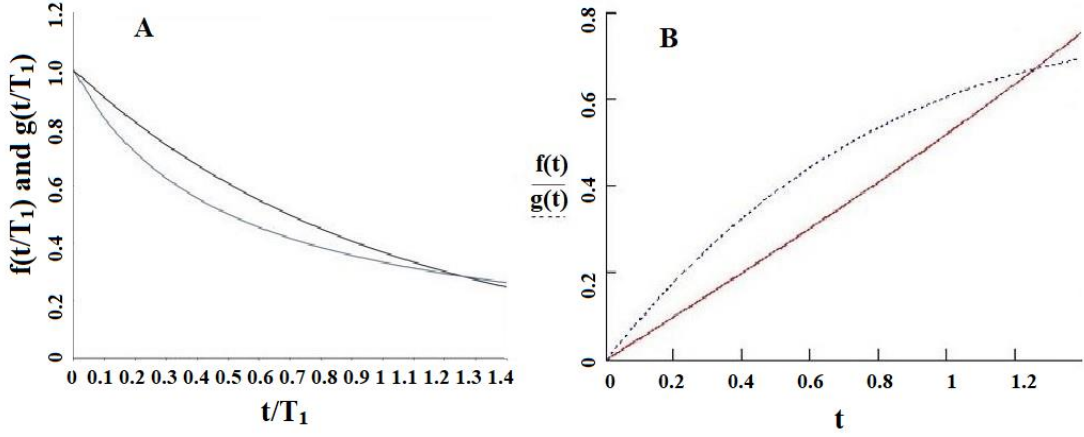

**Figure 1**. (A) Graphical solution of equation (13) with $f\left(\frac{t}{T_1}\right) = e^{-\frac{t}{T_1}}$ and $g\left(\frac{t}{T_1}\right) = \frac{T_1}{2t+T_1}$, (B) Graphical solution of equation (15) with $f(t) = \sinh\left(\frac{t}{2T_1}\right)$ and $g(t) = t/T_1\left(e^{-\frac{t}{2T_1}}\right)$, with t being the repetition time TR, and $T_1$ the longitudinal relaxation time.

*$T_1$ estimation for three vials of the manganese chloride (MnCl$_2$) phantom.*

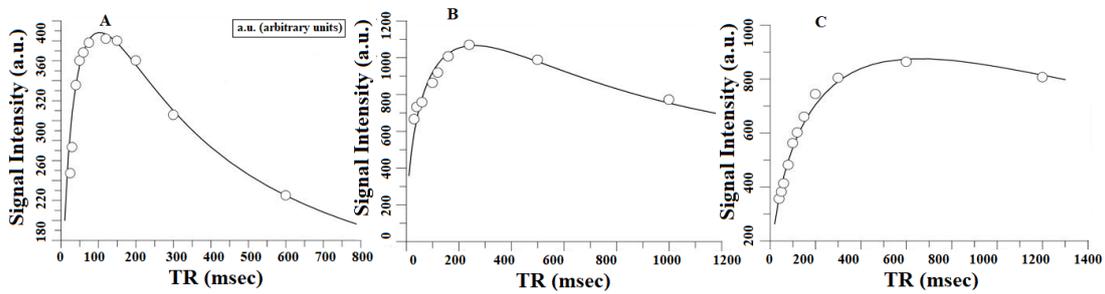

**Figure 2**. Least squares fit with equation (5) of signal to noise ratio (SNR) vs repetition time (TR) data pertaining to the multi-vial study. $T_1$ values shown are (A) $T_1$ = 98.0 ms

with TR ranging between 25 – 600 ms, (B) $T_1$=213.0 ms with TR ranging between 20 and 1000 ms and (C) $T_1$=554.0 ms with TR ranging between 20 – 1260 ms.

*Statistical analyses of the data acquired on the manganese chloride ($MnCl_2$) phantom.*

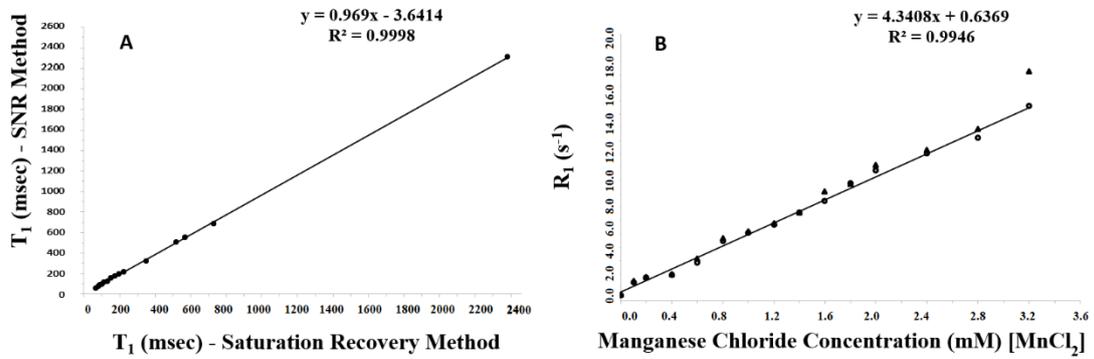

**Figure 3.** (A) Regression plot with $R^2$ correlation constant and (B) comparison of the data corresponding to the manganese (Mn) phantom measurements with the proposed method (triangles) and with conventional saturation recovery pulse sequences (circles), as a function of manganese chloride ($MnCl_2$).

*In vivo acquisitions at two field strengths, in agreement with theoretical analyses.*

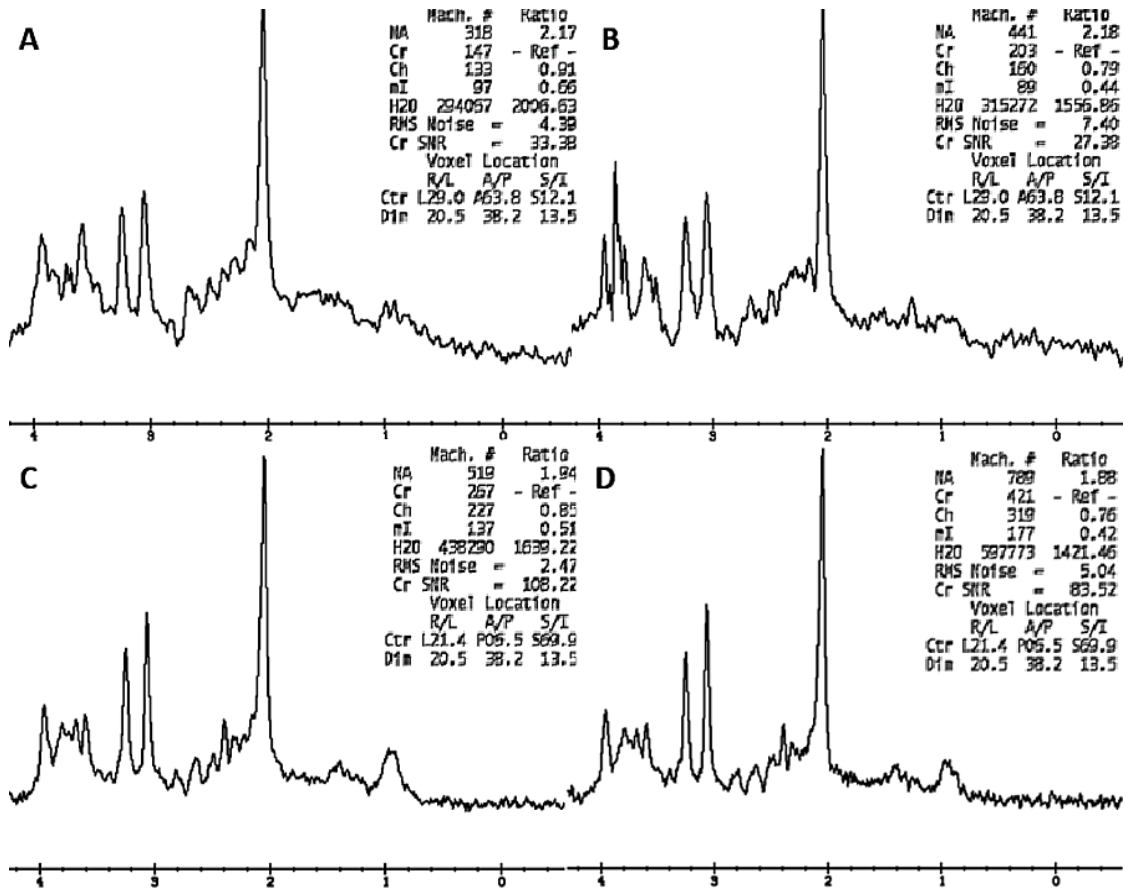

**Figure 4.** Proton magnetic resonance spectra acquired at 1.5T (A, B) and at 3.0T (C, D) from a 10.57 cm$^3$ (1.35 x 2.05 x 3.82 cm$^3$) single voxel placed at the left central semiovale white matter. (A, C): TR=1.5 s, 64 averages and (B, D): TR=6.0 s, 16 averages. TR: repetition time, TE: echo time, NA: N-Acetyl-L-aspartate, Cr: creatine, Ch: choline, mI: myoinositol, RMS: root mean square, SNR: signal to noise ratio.